\documentclass{article}
\usepackage[a4paper, top=35truemm, bottom=30truemm, left=24truemm, right=24truemm]{geometry}
\usepackage{graphicx}
\usepackage{xcolor}
\usepackage{amsmath,amssymb}
\usepackage{amscd}
\usepackage{mathrsfs}
\usepackage{tensor}
\usepackage{mathtools}
\usepackage{enumitem}
\usepackage{authblk}
\usepackage{pdfpages}
\usepackage{cite}
\usepackage{latexsym}
\usepackage{fancyhdr}
\usepackage[normalem]{ulem}
\usepackage{ascmac}
\usepackage{bm}

\usepackage{tikz}
\usepackage[compat=1.1.0]{tikz-feynhand}

\usepackage{appendix}

\newcommand{\nn}{\nonumber\\}

\newcommand{\be}{\begin{eqnarray}}
\newcommand{\ee}{\end{eqnarray}}
\newcommand{\Lag}{{\mathcal L}}
\newcommand{\Ham}{{\mathcal H}}
\newcommand{\B}[1]{{\mathbf{#1}}} 
\newcommand{\Ket}[1]{{\left|{#1}\right\rangle}} 
\newcommand{\Bra}[1]{{\left\langle{#1}\right|}} 
\newcommand{\Braket}[2]{{\left\langle{#1} \middle| {#2}\right\rangle}} 
\newcommand{\cO}[1]{{\cal O}\left({#1} \right)} 
\newcommand{\M}{{\cal M}} 
\newcommand{\Ima}{{\mbox{Im}\,}}

\begin{document}

\title{Perturbative $S$-matrix Unitarity \\ in Higher-Derivative Theory of Fermions}
\author[1]{Yugo Abe}
\author[2,3]{Takeo Inami}
\author[4,5]{Keisuke Izumi}
\author[6]{Shinsuke Kawai}

\affil[1]{Department of Pure and Applied Physics, Kansai University, Suita 564-8680, Japan}
\affil[2]{iTHEMS Program, RIKEN, Wako 351-0198, Japan}
\affil[3]{Yukawa Institute for Theoretical Physics, Kyoto University, Kyoto 606-8502, Japan}
\affil[4]{Department of Mathematics, Nagoya University, Nagoya 464-8602, Japan}
\affil[5]{Kobayashi-Maskawa Institute, Nagoya University, Nagoya 464-8602, Japan}
\affil[6]{School of Science, Yamagata University, Yamagata, 990-8560, Japan}

\date{}

\maketitle
\vspace{-10mm}

\begin{abstract}
We construct a higher-derivative fermion theory with four-Fermi interactions. 
In quantum field theory, it is conjectured that good high-energy behavior of tree-level scattering amplitudes implies renormalizability, a notion referred to as the tree-unitarity criterion. 
To investigate whether this criterion is satisfied in a higher-derivative fermion theory, we examine the structure of the $S$-matrix in detail. 
Owing to the higher-derivative kinetic term, the mass dimension of the spinor field differs from that of a conventional fermion, rendering four-Fermi interactions renormalizable at the cost of introducing ghost states. 
We show that, unlike bosonic ghosts, fermionic ghosts inevitably give rise to negative-norm states regardless of the sign of their energy eigenvalues. 
We further demonstrate that, in contrast to conventional four-Fermi theories, the higher-derivative four-Fermi theory satisfies perturbative $S$-matrix unitarity. 
In light of the tree-unitarity criterion, these results provide supporting evidence for the renormalizability of higher-derivative fermion theories and generalize our previous findings on higher-derivative scalar theories to the fermionic case.
\end{abstract}


\section{Introduction}
Whether quantum gravity plays a role in particle physics is a tantalizing question. 
Such a question is meaningful only if gravity admits a renormalizable quantum description. 
Among the most extensively studied approaches to renormalizable quantum gravity are string field theory and quadratic gravity~\cite{Stelle:1976gc}. 
Since closed string field theory has not yet been fully established, quadratic gravity, characterized by a higher-derivative (HD) kinetic term for the graviton and known to be renormalizable, remains a promising framework. 
A well-known difficulty of quadratic gravity is that, depending on the quantization prescription, it contains either states with energy unbounded from below or negative-norm states (ghosts). 
This feature has been studied extensively since the pioneering work of Lee and Wick~\cite{Lee:1970iw}; see~\cite{Salvio:2018crh} for a recent review. 
To gain a deeper understanding of the subtle quantum behavior of ghosts, HD theories of lower-spin fields, including scalar fields~\cite{Lee:1969fy, Hawking:2001yt, Abe:2018rwb} and vector fields~\cite{Lee:1970iw}, have been investigated in detail. 
In contrast, HD fermion theories remain comparatively underdeveloped, with only partial results available~\cite{Lee:1970iw, Ferrara:1977mv, Kaku:1978nz, Fujimori:2017kyi, Fujimori:2016udq}.

To study the ultraviolet (UV) consistency of quantum field theories, scattering amplitudes serve as a useful probe. 
In such studies, {\it unitarity} plays a key role, as emphasized in recent investigations including \cite{Adams:2006sv,Tokuda:2020mlf,Alberte:2020jsk}. 
For example, the celebrated bound on total cross sections due to Froissart \cite{Froissart:1961ux} and Martin \cite{Martin:1962rt} results from unitarity of the scattering amplitudes. 
It is also anticipated that unitarity of the scattering amplitudes is closely related to the renormalizability of the theory; this idea dates back to the 1970s \cite{Bell:1973ex,LlewellynSmith:1973yud}. 
In particular, in many theories, the conditions for perturbative renormalizability \cite{tHooft:1971qjg} are observed to coincide with unitarity at the tree level \cite{Cornwall:1973tb,Cornwall:1974km}. 
The validity of this tree-unitarity criterion for renormalizability has also been checked for non-relativistic Lifshitz-type scalar models in \cite{Fujimori:2015wda,Fujimori:2015mea}.

The concept of unitarity in fact consists of two separate conditions. 
One is $S$-matrix unitarity, namely $SS^\dagger=1$ for the scattering matrix $S$, which means that the time evolution of states is unique. 
The other is the positive norm of the states, which makes the probability positive.
Since quadratic gravity is known to be perturbatively renormalizable \cite{Stelle:1976gc} despite being formulated as an HD theory that involves negative-norm states, it is interesting to ask whether the aforementioned connection between renormalizability and tree-unitarity can persist without norm positivity. 
As the traditional analysis assumes both $S$-matrix unitarity and norm positivity, and hence is not applicable to HD theories, in Ref.~\cite{Abe:2018rwb} we reinvestigated the analysis from the optical theorem and reformulated the framework without assuming norm positivity. 
Focusing on tree-level amplitudes and not assuming the positivity of norms, we showed that an inequality derived from $S$-matrix unitarity alone is equivalent to the condition for renormalizability in the case of HD scalars. 
This correspondence was later extended to the graviton-matter sector of quadratic gravity~\cite{Abe:2022spe}. 
These examples indicate that $S$-matrix unitarity of tree-level amplitudes at high energies is a powerful criterion for ultraviolet consistency. 
We refer to this criterion as $S$-matrix tree-unitarity.

In the present paper, we apply the framework of $S$-matrix tree-unitarity to the theory of HD fermions. 
We first formulate the theory of HD fermions and describe the formalism of canonical quantization\footnote{ 
It is known that the standard (positive norm) Dirac theory can be reformulated as a second-derivative theory \cite{Fock:1937dy,Feynman:1958ty,Brown:1958zz,Chalmers:1997ui,Espin:2013wia,Espin:2015bja}. 
Our theory is distinct from such reformulations as ours involves negative-norm fermions. 
In this sense, it can be regarded as the closest fermionic cousin of the HD fields relevant to the study of quadratic gravity. 
}. 
We incorporate quartic self-interactions of the HD fermion fields, which turn out to be renormalizable, unlike the case of standard four-Fermi interactions. 
By evaluating the scattering amplitudes for particle-particle processes, we then demonstrate that the $S$-matrix tree-unitarity criterion is satisfied for the HD fermions.

The study of negative-norm states (ghosts) arising from HD kinetic terms has a long history \cite{Dirac:1942,Froissart:1959occ,Lee:1969fy}. 
The possible violation of unitarity induced by such ghosts has been extensively discussed in the literature, particularly from the perspective of complex poles in the energy plane \cite{Lee:1969fy,Nakanishi:1971jj,Donoghue:2023yjt}; see Ref.~\cite{Salvio:2018crh} for a recent review and further references. 
This issue was revisited more recently by Kugo and Kubo \cite{Kubo:2023lpz,Kubo:2024ysu} from the standpoint of the operator formalism.
In this paper, we take the position that negative-norm states are not necessarily fatal to the consistency of a theory. 
Indeed, it is interesting to note that, although negative-norm states are traditionally regarded as a threat to unitarity, our analysis reveals that, at least from the perspective of tree-level $S$-matrix unitarity, they are indispensable and appear to play an essential role in ensuring the ultraviolet consistency of HD theories.

We adopt the following conventions.
Greek indices ($\mu, \nu, \dots$) denote four-dimensional spacetime components, ranging from $0$ to $3$.
Latin indices from the middle of the alphabet ($i, j, \dots$) refer to three-dimensional spatial components ($1, 2, 3$).
Spinor indices are represented by lowercase Greek letters from the beginning of the alphabet ($\alpha, \beta, \dots$) and run from $1$ to $4$ for Dirac spinors.
Bold letters $\B x, \B y, \dots$ and $\B p, \B q, \dots $ represent three-dimensional position coordinates and momenta, respectively.
The metric signature is $\eta_{\mu\nu} = \text{diag}(1, -1, -1, -1)$ and
the Dirac matrices are defined such that $\gamma^0$ is Hermitian, while $\gamma^i$ are anti-Hermitian.

The remainder of the paper is organized as follows. 
In Section~\ref{Sec:Dec}, we describe our theory of higher derivative fermions and in Section~\ref{sec:canonical_quantization} we discuss canonical quantization.
Section~\ref{Sec:TL} presents our main results on the computation of the scattering amplitude and demonstrates that the theory satisfies S-matrix tree-unitarity.
In Section~\ref{Sec:Sum} we conclude with brief comments.
Appendix~\ref{ApFeynRule} presents the Feynmann rules for our theory. 
In Appendix~\ref{ApCal}, the detailed calculation of the scattering amplitudes is shown.


\section{Decomposition of the higher derivative kinetic term} \label{Sec:Dec}

We introduce the higher-derivative extension of fermions.
Since the canonical fermion fields feature first-order derivatives in the Lagrangian,
the simplest HD extension involves second-order derivatives.
We focus on this case in the present paper.
We thus start with a free Lagrangian, 
\begin{eqnarray}
\Lag_2 (\Psi,\bar\Psi) = - \bar{\Psi} (i \gamma^\mu \partial_\mu - m_1) (i \gamma^\nu \partial_\nu - m_2) \Psi,
\label{HD2Lag}
\end{eqnarray}
where $\Psi$ is a 4-component spinor field and $\bar{\Psi}$ is its Dirac conjugate. 
We assume the two unequal mass parameters $0< m_1 < m_2$.
Since the kinetic term contains second-order derivatives, this theory is expected to describe twice as many degrees of freedom as the canonical case. 
It is therefore useful to reformulate this theory in terms of two distinct spinor fields.

To calculate the scattering amplitude, we need to know how interaction terms are transformed under the decomposition of the fermion into two spinor fields. 
We consider a generic interaction term $\mathcal{L}_{\text{int}}(\Psi, \bar{\Psi})$ and examine its transformation. 
The total Lagrangian is
\begin{eqnarray}
\Lag = \Lag_2(\Psi,\bar\Psi) + \Lag_{\mbox{int}}(\Psi,\bar\Psi) .
\label{HDLagWint}
\end{eqnarray}

We replace the derivative term $i\gamma^\mu \partial_\mu \Psi$ with an auxiliary field $\zeta$. This is achieved by introducing the following constraint term with a Lagrange multiplier $\bar{\chi}$:
\begin{eqnarray}
\bar\chi\left( i\gamma^\mu\partial_\mu \Psi -\zeta \right). \label{multi}
\end{eqnarray}
We also add the Hermitian conjugate of \eqref{multi} in order to ensure the reality of the Lagrangian. 
The Lagrangian then reads
\begin{eqnarray}
&&\Lag =- \frac12\left[
\overline{\left( \zeta-m_1\Psi \right)} \left( \zeta-m_2\Psi \right) +\overline{\left( \zeta-m_2\Psi \right)} \left( \zeta-m_1\Psi \right) 
\right. \nn 
&& \qquad\qquad\qquad \left.
+2\bar\chi\left( i\gamma^\mu\partial_\mu \Psi -\zeta \right) + 2\overline{\left( i\gamma^\mu\partial_\mu \Psi -\zeta \right)}\chi
\right] 
+ \Lag_{\mbox{int}}(\Psi,\bar\Psi) .
\label{LagPz}
\end{eqnarray}
A straightforward calculation shows that the Lagrangian \eqref{LagPz} becomes
\begin{eqnarray}
&&\Lag = - \overline{\left( \zeta-\frac12 (m_1+m_2)\Psi - \chi \right)} \left( \zeta-\frac12 (m_1+m_2)\Psi - \chi \right) - i \bar \chi \gamma^\mu \partial_\mu \Psi + i \left(\partial_\mu \bar \Psi\right) \gamma^\mu \chi\nn
&&\qquad \qquad   + \frac14(m_1-m_2)^2 \bar \Psi \Psi + \frac12(m_1+m_2)(\bar\Psi \chi + \bar \chi \Psi) + \bar \chi \chi + \Lag_{\mbox{int}}(\Psi,\bar\Psi) .
\end{eqnarray}
The first term on the right-hand side can be eliminated by substituting the constraint equation obtained by variation with respect to $\bar \zeta$:
\begin{eqnarray}
\zeta-\frac12 (m_1+m_2)\Psi - \chi =0.
\end{eqnarray}
Then the Lagrangian becomes independent of $\zeta$ and depends only on $\Psi$ and $\chi$.
Under the field redefinition
\begin{eqnarray}
\Psi &=& \frac{-1}{\sqrt{m_2-m_1}} \left(\psi_1 - \psi_2\right), \label{FTPsi} \\
\chi &=& \frac{\sqrt{m_2-m_1}}{2}\left(\psi_1 + \psi_2 \right),
\end{eqnarray}
we finally obtain the Lagrangian in terms of the two spinor fields $\psi_1$ and $\psi_2$:
\begin{eqnarray}
\Lag = \bar \psi_1 (i \gamma^\mu \partial_\mu - m_1) \psi_1 - \bar \psi_2 (i \gamma^\mu \partial_\mu - m_2) \psi_2 + \Lag_{\mbox{int}}\left(\frac{-\psi_1 + \psi_2}{\sqrt{m_2-m_1}}, \frac{-\bar\psi_1 + \bar\psi_2}{\sqrt{m_2-m_1}}\right).
\label{Lag12}
\end{eqnarray}

We make two remarks on the Lagrangian \eqref{Lag12}. 
First, the kinetic term of $\psi_2$ has the opposite sign from that of $\psi_1$ which is canonical. 
This feature is generic when an HD theory is reformulated in terms of multiple non-HD fields. 
In the next section, we will see that this sign-flipped kinetic term leads to the existence of negative-norm states; a field with this property is often referred to as a ghost field.
Second, this transformation does not alter the functional form of the interaction Lagrangian. 
Hence, the interaction terms in the new basis are obtained simply by substituting the field transformation \eqref{FTPsi} into $\Lag_{\text{int}}$.

\section{Canonical quantization}\label{sec:canonical_quantization}
In this section, we examine the quantization of the ghost fermion $\psi_2$. 
In the canonical quantization of a ghost field, such as $\psi_2$ in the Lagrangian~(\ref{Lag12}), the overall minus sign introduces a crucial difference from the standard case and gives rise to negative-norm states. 
To clarify its origin, we present the procedure of canonical quantization in detail. 
We also show that negative-norm states are unavoidable for ghost fermions, in contrast to the case of scalar ghost fields. 
For scalar ghost fields, an alternative choice exists, in which negative-norm states are avoided at the cost of negative-energy modes.\footnote{
Indefinite-metric formulations of fermionic theories have a long history; see, for example, Heisenberg's discussion of his nonlinear spinor theory, including the relation between an indefinite metric and the unitarity of the $S$-matrix~\cite{Heisenberg:1957zz}.}

We consider the free part of the first-derivative Lagrangian \eqref{Lag12} and, for illustrative purposes, treat the canonical and ghost fields collectively as
\begin{eqnarray}
\Lag = \pm \bar \psi (i \gamma^\mu \partial_\mu - m) \psi.
\label{Lagpsi}
\end{eqnarray}
The plus ($+$) case in the overall sign $\pm$ corresponds to the canonical fermion $\psi_1$, while the minus ($-$) case represents the ghost fermion $\psi_2$.
The canonical momentum is obtained by the right variation with respect to $\dot \psi_\alpha=\partial_{0}\psi_{\alpha}$ as
\begin{eqnarray}
\pi^\alpha := \frac {\partial_R \Lag}{\partial \dot \psi_\alpha} = \pm i (\bar \psi \gamma^0)^\alpha = \pm i \psi^{\dagger\alpha},
\end{eqnarray}
where $\partial_R \Lag / \partial \dot \psi_\alpha$ represents the right derivative of $\Lag$ with respect to $\dot \psi_\alpha$.  
We introduce the anti-commutation relations for the canonical variable $\psi_\alpha$ and its conjugate momentum $\pi^\alpha$:
\begin{eqnarray}
&&\{\psi_\alpha (\B x), \pi^\beta(\B y)\} = i \delta^\beta_\alpha \delta^3(\B x  - \B y),  \\
&&\{\psi_\alpha (\B x), \psi_\beta(\B y)\}=\{\pi_\alpha (\B x), \pi_\beta(\B y)\} =0,
\end{eqnarray}
which lead to
\begin{eqnarray}
&&\{\psi_\alpha (\B x), \psi^{\dagger\beta}(\B y)\} = \pm \delta^\beta_\alpha \delta^3(\B x  - \B y), \label{psipi} \\
&&\{\psi_\alpha (\B x), \psi_\beta(\B y)\}=\{\psi^{\dagger\alpha} (\B x), \psi^{\dagger\beta}(\B y)\} =0.
\end{eqnarray}
Corresponding to the overall sign of the Lagrangian \eqref{Lagpsi}, the sign difference appears in the anti-commutation relation \eqref{psipi}.

The Hamiltonian $\Ham$ is defined as
\begin{eqnarray}
\Ham &:=& \int d^3 \B x  \left[ \pi^\alpha \dot \psi_\alpha - \Lag \right] \nn
&=& \int d^3 \B x  \left[\pm i \psi^{\dagger\alpha} \dot \psi_\alpha \mp \bar \psi (i\gamma^\mu \partial_\mu - m)\psi \right]\nn
&=& \int d^3 \B x  \left[  \mp \bar \psi (i\gamma^k \partial_k - m)\psi \right].
\end{eqnarray}
The Heisenberg equation is given by
\begin{eqnarray}
i\dot \psi_\alpha &=& [\psi_\alpha , \Ham ] \nn
&=& - \gamma^0 (i\gamma^k \partial_k - m)\psi_\alpha,
\end{eqnarray}
which is equivalent to the Euler-Lagrange equation
\begin{eqnarray}
(i\gamma^\mu\partial_\mu - m ) \psi_\alpha=0
\end{eqnarray}
derived from the Lagrangian \eqref{Lagpsi}. 
It is important to note that the Heisenberg equation remains the same for both the canonical and ghost fields. Consequently, the solutions to the Heisenberg equation are obtained in the canonical form
\begin{eqnarray}
\psi (\B x)= \int \frac{d^3 p}{\sqrt{(2\pi)^3 2 p_0}} \sum_{s=\pm}
\left[
a(\B p,s) u(\B p,s) \exp(-ip_\mu x^\mu)+b^{\dagger}(\B p,s) v(\B p,s) \exp(ip_\mu x^\mu)
\right],
\end{eqnarray}
where
\begin{eqnarray}
p_0:= \sqrt{\B p^2+m^2}.
\end{eqnarray}
The spinor bases are normalized as
\begin{eqnarray}
\bar u (\B p,s) u(\B p,s')=2m \delta_{ss'}, \quad \bar v (\B p,s) v(\B p,s')=-2m \delta_{ss'}.
\end{eqnarray}
The anti-commutation relations for the operators $a(\B p,s), a^\dagger(\B p,s), b(\B p,s)$, and $b^\dagger(\B p,s)$ are found to be
\begin{eqnarray}
\{ a(\B p,s),a^\dagger(\B q,s')\}= \{ b(\B p,s),b^\dagger(\B q,s')\}=\pm \delta_{ss'}\delta^3(\B p-\B q),
\label{anti-com}
\end{eqnarray}
with all other anticommutators vanishing.
The minus-sign case corresponding to the ghost fermion was considered in \cite{Ferrara:1977mv,Kaku:1978nz}.
Our treatment of the canonical and ghost fermions differs from, and is simpler than, that in these references. 
We therefore present the quantization procedure in full detail.

Due to the overall sign of the Lagrangian \eqref{Lagpsi}, the corresponding sign difference appears in the anti-commutation relations \eqref{anti-com}. 
In contrast to the commutation relations for bosonic fields, the negative sign in the anti-commutators of the ghost field cannot be removed by interchanging the roles of the creation and annihilation operators. 
Since the anti-commutator is symmetric in its arguments, the sign in Eq.~\eqref{anti-com} remains unchanged even when the order of the creation and annihilation operators is reversed. 
Therefore, the existence of negative-norm states is inevitable. 
This constitutes an essential difference from bosonic ghost fields, for which two distinct quantization prescriptions are available~\cite{Abe:2018rwb}. 
One yields negative-norm states with positive energy, whereas the other gives positive-norm states with negative energy. 
For a fermionic ghost, 
interchanging the roles of the creation and annihilation operators
merely changes the energy of the one-particle state from positive to negative without affecting the sign of its norm. 
Therefore, as in the case of an ordinary canonical field, we adopt the convention in which the physical excitations have positive energy. 
This choice corresponds to the conventional interpretation in terms of the Dirac sea.

Returning to the Lagrangian \eqref{Lag12}, the mode expansion for $\psi_I$ $(I=1,2)$ is expressed as
\begin{eqnarray}
\psi_I (\B x)= \int \frac{d^3 p}{\sqrt{(2\pi)^3 2 p_0}} \sum_{s=\pm}\left[a_I(\B p,s) u_I(\B p,s) \exp(-ip_\mu x^\mu)+b_I^{\dagger}(\B p,s) v_I(\B p,s) \exp(ip_\mu x^\mu)\right].
\end{eqnarray}
The anti-commutation relations for the creation and annihilation operators are given by
\begin{eqnarray}
\{ a_I(\B p,s),a_J^\dagger(\B q,s')\}
= \{ b_I(\B p,s),b_J^\dagger(\B q,s')\}
= \epsilon_I\delta_{IJ} \delta_{ss'}\delta^3(\B p-\B q), \quad \mbox{with} \quad \epsilon_1=1, \quad \epsilon_2=-1
\end{eqnarray}
and all other anti-commutators vanish. The vacuum state $\Ket{0}$ is defined by
\begin{eqnarray}a_I(\B p,s) \Ket{0}=b_I(\B p,s) \Ket{0}=0.\end{eqnarray}
The Fock space is constructed as usual by acting with the creation operators $a_I^\dagger(\B p,s)$ and $b_I^\dagger(\B p,s)$ on the vacuum state.
For later convenience, we denote a particle and an antiparticle associated with the field $\psi_I$ by $F_I$ and $\overline{F}_I$, respectively.
Hence, the one-particle states of $F_I$ and $\overline{F}_I$ are defined as
\begin{eqnarray}
\Ket{F_I,\B p,s} := a_I^\dagger(\B p,s)\Ket0,
\qquad
\Ket{\overline{F}_I,\B p,s} := b_I^\dagger(\B p,s)\Ket0.
\end{eqnarray}
Here, the bar on $\overline{F}_I$ denotes an antiparticle and should not be confused with the Dirac conjugation of the spinor field $\bar\psi_I$.
The norm of a one-particle ghost state, for example $\Ket{F_2,\B p,s}$, is given by
\begin{eqnarray}
\Braket{F_2,\B p,s}{F_2,\B p,s}&=& \Bra0 a_2(\B p,s) a_2^\dagger(\B p,s) \Ket0 = \Bra0 \left( \{a_2(\B p,s),a_2^\dagger(\B p,s)\} - a_2^\dagger(\B p,s)a_2(\B p,s) \right) \Ket0 \nonumber\\
&=& -\delta_{ss'}\delta^3(0) < 0,
\end{eqnarray}
showing that a one-particle state of the ghost fermion has a negative norm.


\section{Tree-level four-point scattering amplitude}\label{Sec:TL}

We evaluate fermion-fermion scattering amplitudes. 
Historically, the four-Fermi coupling was used to describe the muon decay $\mu \to e + \bar{\nu}_e + \nu_\mu$, the neutrino scattering $\nu_\mu + e \to \nu_\mu + e$, and other weak interaction processes. 
The focus of our study, however, is different. 
Our aim in constructing the HD fermion theory is to investigate the $S$-matrix unitarity properties of fermion scattering processes, $F_I + F_J \to F_K + F_L$, at high energies. 
The application of our new fermion model to weak interaction processes is currently not our intention.

Let us now return to the Lagrangian \eqref{HDLagWint} for $\Psi$. 
Since the kinetic term $\bar\Psi (i\gamma^\mu\partial_\mu)(i\gamma^\nu\partial_\nu)\Psi$ implies that the mass dimensions of the fields $\Psi$ and $\bar\Psi$ are both unity, the four-Fermi interactions for $\Psi$ and $\bar\Psi$ are power-counting renormalizable. 
We may think of five types of such four-Fermi interactions: the scalar-type, the pseudo-scalar-type, the vector-type, the axial-vector-type and the tensor-type.
Below we focus on the scalar type, as the other four types are expected to be similar so far as the S-matrix unitarity is concerned.
The scalar-type four-Fermi interaction is given by
\begin{eqnarray}
&&\Lag_s := \frac{\lambda_s}{2} (\bar \Psi \Psi)^2= \frac{\lambda_s}{2(m_2-m_1)^2} \left(\sum_{I,J} \epsilon_{I}\epsilon_{J} \bar\psi_I \psi_J\right)^2.
\label{Lags}
\end{eqnarray}
The kinetic terms of $\psi_I$ contain only first-order derivatives, and thus the mass dimensions of $\psi_I$ and $\bar{\psi}_I$ are $3/2$. 
This indicates that the four-Fermi couplings for $\psi_I$ are dimension-six operators, which are not power-counting renormalizable. 
This feature is seen in the interaction Lagrangian, Eq.~\eqref{Lags}, where the coupling constant $\lambda_s/(m_2-m_1)^2$ has a mass dimension of $-2$.
If unitarity and renormalizability are related, the non-renormalizable interaction \eqref{Lags} implies a violation of unitarity.
We examine this correspondence by explicit computations below.

Since unitarity assumes both $S$-matrix unitarity ($SS^\dagger = 1$) and positive-definite norms, the conventional discussions of unitarity bounds \cite{Cornwall:1973tb,Cornwall:1974km} do not apply to HD theories containing negative-norm states. 
In our previous paper~\cite{Abe:2018rwb}, we proposed a generalized version of the unitarity bounds that does not require norm positivity. 
Using that formalism, we investigated HD scalar theories and demonstrated that the condition for renormalizability is equivalent to perturbative $S$-matrix unitarity at tree level. 
We referred to the latter condition as $S$-matrix tree unitarity. 
The aim of this section is to examine whether $S$-matrix tree unitarity is satisfied in a theory of HD fermions by explicitly computing scattering amplitudes arising from a scalar-type four-Fermi interaction. 
We begin with a brief review of $S$-matrix tree unitarity~\cite{Abe:2018rwb}.

$S$-matrix unitarity is expressed as
\begin{eqnarray}\label{eqn:SSdag1}
SS^\dagger=1,
\end{eqnarray}
which, in terms of the transition amplitude $T$ defined by $S=1+iT$, becomes
\begin{eqnarray}
-i(T-T^\dagger)=T\,T^\dagger. \label{TT}
\end{eqnarray}
We introduce a complete orthonormal basis $\{\Ket{X}\}$ that satisfies
\begin{eqnarray}
\sum_X \Ket{X}C_X\Bra{X} =1 \qquad \mbox{with} \qquad \Braket{X}{Y}=C_X^{-1}\delta_{XY},
\end{eqnarray}
where $\Braket{X}{X}\neq 0$ is assumed. 
Note that $\Braket{X}{X}$ can be negative since negative-norm states are allowed. 
Let us define scattering amplitude $\M(i \to f)$ as
\begin{eqnarray}
\Bra{f}T\Ket{i}= \delta^4(p_i-p_f) \M(i \to f),
\end{eqnarray}
where $p_i$ and $p_f$ are the total 4-momenta of the states $\Ket{i}$ and $\Ket{f}$, respectively. 
Eq.~\eqref{TT} is expressed in terms of the scattering amplitude $\M(i \to f)$ as
\begin{eqnarray}
-i[\M(i \to f) -\M(f \to i)^*] = \sum_X C_X \delta^4(p-p_X)\M(i \to X)\M(f \to X)^* ,
\end{eqnarray}
where $p\,(=p_i=p_f)$ is the total 4-momentum, which is conserved in the scattering process. For forward scattering, $\Ket{i}=\Ket{f}$, the above equation becomes
\begin{eqnarray}
2\Ima \M(i\to i) = \sum_X C_X \delta^4(p_i-p_X) |\M(i\to X)|^2 .
\end{eqnarray}
We also have an algebraically trivial inequality
$\left|\M(i\to i) \right| \ge \left| \Ima \M(i\to i) \right|$, 
which leads to
\begin{eqnarray}
\left|\M(i\to i) \right| \ge \left| \Ima \M(i\to i) \right| = \frac12 \left| \sum_X C_X \delta^4(p_i-p_X) |\M(i\to X)|^2 \right| .
\label{UB}
\end{eqnarray}

Let us first comment on the standard procedure for obtaining bounds from the optical theorem when negative-norm states are absent. 
In this case, it is convenient to use a discrete basis of states for which all $C_X=1$. 
Since each term in the sum on the right-hand side of \eqref{UB} is positive, a partial sum does not exceed the total sum. 
Taking the partial sum over the on-shell states that constitute a subset of $X$, one obtains inequalities that are used to derive the Froissart–Martin bound \cite{Froissart:1961ux,Martin:1962rt} and various unitarity bounds \cite{LlewellynSmith:1973yud,Cornwall:1973tb,Cornwall:1974km,Lee:1977eg,Fujimori:2015mea}.

When negative-norm states are present, in contrast, $C_X$ can be negative, so that a partial sum can even exceed the total sum, and the above argument no longer applies.
We therefore evaluate the inequality \eqref{UB} as it is, using a continuous basis of states rather than a discrete basis. 
The sum in \eqref{UB} is then understood as an integration over an angle variable $\theta$ (see below). 
The $S$-matrix tree-unitarity criterion thus stands for the inequality \eqref{UB} itself, which is simply the $S$-matrix unitarity condition \eqref{eqn:SSdag1} combined with the optical theorem, evaluated at the tree-level and at the asymptotic energy scale.

We will examine the condition \eqref{UB} in the following subsections \ref{Subsec:TL1} and \ref{Subsec:TL2}. 
The $S$-matrix tree-unitarity is violated for the standard four-Fermi theory with a canonical kinetic term, which is consistent with the fact that it is not power-counting renormalizable. 
In our HD fermion theory with the four-Fermi coupling, elements in the scattering amplitude also exhibit high-energy behavior that may violate the inequality, but the $S$-matrix tree-unitarity criterion is shown to be satisfied since cancellation takes place among the terms of the sum in \eqref{UB}. 
We will demonstrate this at leading order in the coupling constant $\lambda_s$, focusing for this purpose on the $2 \to 2$ scattering processes. 
It may appear that there are many distinct initial states $\Ket{F_I, F_J}$, but the scalar-type four-Fermi interaction \eqref{Lags} is invariant under charge conjugation, parity inversion, and spatial rotations, and many of the resulting amplitudes are not independent. 
We thus consider two representative cases: an initial state consisting of two $F_1$ particles with spins $(+,+)$ and an initial state consisting of two $F_1$ particles with spins $(+,-)$.

We denote the initial state $\Ket{i}$ consisting of two canonical particles $F_1$ with spins $s$ and $s'$ as
\begin{eqnarray}\Ket{i}= \left((2\pi)^{3/2} \sqrt{2E_{\B p,1}}, a_1^\dagger(\B p,s)  \right) \left((2\pi)^{3/2} \sqrt{2E_{-\B p,1}}, a_1^\dagger(-\B p,s')  \right) \Ket{0} = : \Ket{1 ,\B p, s; 1 ,-\B p, s'},\end{eqnarray}
with
\begin{eqnarray}
E_{\B p,I}:= \sqrt{\B p^2+ m_I^2}.
\end{eqnarray}
We work in the center-of-mass frame, 
and the factor $(2\pi)^{3/2} \sqrt{2E_{\B p,1}}$ is introduced to ensure that the creation operators are in a covariant form. 
Similarly, the notation $\Ket{I ,\B p, s; J ,\B q, s'}$ denotes a two-particle state consisting of a $F_I$ particle with momentum $\B p$ and spin $s$,
and a $F_J$ particle with momentum $\B q$ and spin $s'$.

The scattering amplitudes are calculated using the Feynman rules based on these covariant creation operators.
The Feynman rules for the scalar-type four-Fermi couplings are provided in Appendix~\ref{ApFeynRule}.
For the scalar-types four-Fermi interaction in Eq.\eqref{Lags}, 
the possible final states are $\Ket{1 ,\B q, s; 1 ,-\B q, s'}$, $\Ket{1 ,\B q, s; 2 ,-\B q, s'}$, and $\Ket{2 ,\B q, s; 2 ,-\B q, s'}$. 
Therefore, to verify $S$-matrix tree-unitarity, we need to evaluate the transitions to these final states. 
Using the results of calculations in Appendix~\ref{ApCal}, 
we confirm below that the $S$-matrix tree-unitarity is satisfied.
The key feature is the cancellations in the summation on the right-hand side of Eq.~\eqref{UB}.

\subsection{$(+,+)$-spins initial state}\label{Subsec:TL1}

As detailed in Appendix~\ref{ApCal}, the elements of the scattering amplitudes contain $\cO{| \B p|^4}$ contributions, which diverge in the UV limit. 
This behavior violates the conventional unitarity bound. 
However, we will see below, $S$-matrix tree-unitarity is reproduced after the summation over $X$ is taken in Eq.~\eqref{UB}.

The amplitude for the process 
$F_1(\B p, +)F_1(-\B p, +) \to F_1(\B q, +)F_1(-\B q, +)$ 
is given by
\begin{eqnarray}
 \M \bigl( F_1(\B p, +)F_1(-\B p, +) \to F_1(\B q, +)F_1(-\B q, +) \bigr)
= \frac{\lambda_s}{(m_2-m_1)^2} \left[4m_1^2 + \cO{\frac{m^4}{|\B p|^2}} \right].
\end{eqnarray}
Since this is independent of $\theta$, the forward scattering amplitude takes the same value, whose leading order is $\cO{| \B p|^0}$. 
To satisfy the inequality~\eqref{UB}, the right-hand side must be at most of order $\mathcal{O}(|\mathbf{p}|^0)$. 
However, as seen in the result in Appendix~\ref{ApCal}, each individual term in the sum on the right-hand side is of order $\mathcal{O}(|\mathbf{p}|^4)$. 
This implies that cancellation must occur at orders $|\mathbf{p}|^4$ and $|\mathbf{p}|^2$. 
We demonstrate that such cancellations indeed take place. 
At the lowest order in perturbation theory with respect to the coupling constant $\lambda_s$, 
the right-hand side of Eq.~\eqref{UB} becomes (for details of the derivation, see Ref.~\cite{Abe:2018rwb})
\begin{eqnarray}
&&\frac12 \left| \sum_X C_X \delta^4(p_i-p_X) |\M(i\to X)|^2 \right|  \nn
&&\simeq \frac12 \Bigg| \sum_{IJ}\sum_{ss'} \int \frac{d^3 {\mathbf p}_I}{2E_{{\mathbf p}_I,I}} \frac{d^3{\mathbf p}_J}{2E_{{\mathbf p}_J,J}} \delta^4(p_1+p_2-p_I-p_J)(-1)^{n_{IJ}}\nn
&&\hspace{23mm}\times
\Big| \M \bigl( F_1({\mathbf p},+)F_1(-{\mathbf p},+) \to F_I({\mathbf p}_I,s) F_J({\mathbf p}_J,s') \bigr) \Big|^2 \Bigg|.
\end{eqnarray}
where $n_{IJ}$ is the number of ghost particles $F_2$ in the final state, and $p_I$ in the Dirac's delta function denotes the four-momentum corresponding to ${\mathbf p}_I$, that is, $p_I:=(E_{{\mathbf p}_{I},I},{\mathbf p}_{I})$.
Performing the integration over ${\mathbf p}_J$ by means of the spatial delta function, 
we have ${\mathbf p}_I=-{\mathbf p}_J:={\mathbf q}_{IJ}$, {\it i.e.}
the four momenta $p_I$ and $p_J$ are expressed as
\begin{eqnarray}
p_I=(E_{{\mathbf q}_{IJ},I},{\mathbf q}_{IJ}),
\qquad
p_J=(E_{-{\mathbf q}_{IJ},J},-{\mathbf q}_{IJ}).
\end{eqnarray}
The magnitude of ${\mathbf q}_{IJ}$ is determined by energy conservation, 
and depends on both the species of final-state particles $(I,J)$.
The expression is then rewritten as
\begin{eqnarray}
\propto \Bigg| \sum_{IJ}\sum_{ss'} \frac{|{\mathbf q}_{IJ}|}{E_{{\mathbf p},1}+E_{-{\mathbf p},1}}\int_0^\pi \sin\theta\, \Big| \M\bigl( F_1({\mathbf p},+)F_1(-{\mathbf p},+) \to F_I({\mathbf q}_{IJ},s) F_J(-{\mathbf q}_{IJ},s') \bigr) \Big|^2 d\theta \Bigg|,
\end{eqnarray}
where $\theta$ is the scattering angle, defined by $\B p\cdot\B q=|\B p||\B q|\cos{\theta}$.
Using the amplitudes given in Appendix~\ref{ApCal}, we evaluate each term in the sum as follows.

For $(I,J)=(1,1)$,  we obtain
\begin{eqnarray}
&&\frac{(-1)^{n_{11}}|\B q_{11}|}{E_{\B p, 1}+E_{-\B p, 1}}  \int_0^\pi \sin \theta\Big|\M \bigl( F_1(\B p, +)F_1(-\B p, +) \to F_1(\B q_{11}, +)F_1(-\B q_{11}, +) \bigr)\Big|^2 d \theta =  \cO{\alpha^2\frac{m^4}{|\B p|^0}}, \nn\\
&&\frac{(-1)^{n_{11}}|\B q_{11}|}{E_{\B p, 1}+E_{-\B p, 1}}  \int_0^\pi \sin \theta\Big|\M \bigl( F_1(\B p, +)F_1(-\B p, +) \to F_1(\B q_{11}, -)F_1(-\B q_{11}, -) \bigr)\Big|^2 d \theta \nn
&&\qquad
=\frac{16\alpha^2|\B q_{11}|}{E_{\B p, 1}+E_{-\B p, 1}}\int_0^\pi \sin \theta \left( \left(|\B p|^2 + m_1^2\right) + \cO{\frac{m^4}{|\B p|^2}} \right)^2 d\theta\nn
&&\qquad
=\frac{16\alpha^2|\B q_{11}|}{E_{\B p, 1}+E_{-\B p, 1}}\int_0^\pi \sin \theta \left(\left[|\B p|^4 +2m_1^2|\B p|^2 \right]
+\cO{\frac{m^4}{|\B p|^0}}\right) d\theta ,
\end{eqnarray}
with 
$|\B q_{11}|= |\B p|$,
where
\begin{eqnarray}
\alpha:= \frac{\lambda_s}{(m_2-m_1)^2}.
\end{eqnarray}
For $(I,J)=(1,2)=(2,1)$, we have
\begin{eqnarray}
&&\frac{(-1)^{n_{12}}|\B q_{12}|}{E_{\B p, 1}+E_{-\B p, 1}}  
\int_0^\pi \sin \theta\Big|\M \bigl( F_1(\B p, +)F_1(-\B p, +) \to F_1(\B q_{12}, +)F_2(-\B q_{12}, +) \bigr)\Big|^2 d \theta =  \cO{\alpha^2\frac{m^4}{|\B p|^0}}, \nn\\
&&\frac{(-1)^{n_{12}}|\B q_{12}|}{E_{\B p, 1}+E_{-\B p, 1}}  
\int_0^\pi \sin \theta\Big|\M \bigl( F_1(\B p, +)F_1(-\B p, +) \to F_1(\B q_{12}, -)F_2(-\B q_{12}, -) \bigr)\Big|^2 d \theta \nn
&&\qquad
=\frac{-\alpha^2|\B q_{12}|}{E_{\B p, 1}+E_{-\B p, 1}}\int_0^\pi \sin \theta \left(-\left[4|\B p|^2+ \frac12\left(7m_1^2+2m_1m_2-m_2^2\right) \right] + \cO{\frac{m^4}{|\B p|^2}}\right)^2 d\theta\nn
&&\qquad
=\frac{-4\alpha^2|\B q_{12}|}{E_{\B p, 1}+E_{-\B p, 1}}\int_0^\pi \sin \theta \left(\left[  4|\B p|^4 + \left(7m_1^2+2m_1m_2-m_2^2\right)|\B p|^2 \right] 
+\cO{\frac{m^4}{|\B p|^0}}\right) d\theta,
\end{eqnarray}
with
\begin{eqnarray}
|\B q_{12}|= |\B p|\left[1 + \frac{m_1^2-m_2^2}{4|\B p|^2} + \cO{\frac{m^4}{|\B p|^4}} \right].
\end{eqnarray}
For $(I,J)=(2,2)$, we have
\begin{eqnarray}
&&\frac{(-1)^{n_{22}}|\B q_{22}|}{E_{\B p, 1}+E_{-\B p, 1}} 
\int_0^\pi \sin \theta\Big|\M \bigl( F_1(\B p, +)F_1(-\B p, +) \to F_2(\B q_{22}, +)F_2(-\B q_{22}, +) \bigr)\Big|^2 d \theta =  \cO{\alpha^2\frac{m^4}{|\B p|^0}}, \nn\\
&&\frac{(-1)^{n_{22}}|\B q_{22}|}{E_{\B p, 1}+E_{-\B p, 1}} 
\int_0^\pi \sin \theta\Big|\M \bigl( F_1(\B p, +)F_1(-\B p, +) \to F_2(\B q_{22}, -)F_2(-\B q_{22}, -) \bigr)\Big|^2 d \theta \nn
&&\qquad
=\frac{\alpha^2|\B q_{22}|}{E_{\B p, 1}+E_{-\B p, 1}}\int_0^\pi \sin \theta \left(\left[4|\B p|^2 +3m_1^2+2m_1m_2-m_2^2\right]
+\cO{\frac{m^4}{|\B p|^2}}\right)^2 d\theta\nn
&&\qquad
=\frac{4\alpha^2|\B q_{22}|}{E_{\B p, 1}+E_{-\B p, 1}}\int_0^\pi \sin \theta \left(\left[4|\B p|^4 + 2\left(3m_1^2+2m_1m_2-m_2^2\right)|\B p|^2\right]
+\cO{\frac{m^4}{|\B p|^0}}\right) d\theta ,
\end{eqnarray}
with
\begin{eqnarray}
|\B q_{22}|= |\B p|\left[1 + \frac{m_1^2-m_2^2}{2|\B p|^2} + \cO{\frac{m^4}{|\B p|^4}} \right].
\end{eqnarray}

Summing these contributions, 
we obtain
\begin{eqnarray}
&&\frac12 \Bigg| \sum_{IJ}\sum_{ss'} \int \frac{d^3p_I}{2E_{\mathbf p_I,I}} \frac{d^3p_J}{2E_{\mathbf p_J,J}} \delta^4(p_1+p_2-p_I-p_J)(-1)^{n_{IJ}}\nn
&&\hspace{12mm}\times
\Big| \M \bigl( F_1(\mathbf p,+)F_1(-\mathbf p,+) \to F_I(\mathbf p_I,s) F_J(\mathbf p_J,s') \bigr) \Big|^2 \Bigg|
= \cO{\alpha^2\frac{m^4}{|\B p|^0}}= \cO{\lambda_s^2|\B p|^0}.
\end{eqnarray}
Noting that for the $F_1 F_2$ state a factor of two is required to account for the two possible configurations: 
one where $F_1$ has momentum $\B q_{12}$ and the other where it has $\B q_{21}=-\B q_{12}$.
In the UV limit, this expression behaves as $\cO{|\B p|^0}$, 
which is consistent with the order of the left-hand side of inequality~\eqref{UB}. 
Furthermore, the left-hand side is of order $\lambda_s$, whereas the right-hand side is of order $\lambda_s^2$.
Therefore, the inequality~\eqref{UB}, and hence $S$-matrix tree-unitarity, is satisfied for sufficiently small couplings where perturbative analysis is valid.

\subsection{$(+,-)$-spins in initial state}\label{Subsec:TL2}
We next consider the case where the initial state has $(+, -)$ spins. 
The scattering amplitude for the process 
$F_1(\B p, +)F_1(-\B p, -) \to F_1(\B q_{11}, +)F_1(-\B q_{11}, -)$ 
is calculated as Eq.~\eqref{+-+-} in Appendix~\ref{ApCal}.
The forward scattering amplitude becomes
\begin{eqnarray}
\M \bigl( F_1(\B p, +)F_1(-\B p, -) \to F_1(\B p, +)F_1(-\B p, -) \bigr)
= \frac{-4\lambda_s}{(m_2-m_1)^2} \left[|\B p|^2 + \cO{\frac{m^4}{|\B p|^2}} \right].
\end{eqnarray}
Since several terms on the right hand side of the inequality~\eqref{UB} contain $\cO{|\B p|^4}$ contributions,
for the S-matrix tree unitarity, all of these contributions must cancel.
If all of $\cO{|\B p|^4}$ contributions cancel out, the right hand side of the inequality~\eqref{UB} becomes $\cO{|\B p|^2}$. 
Then, the inequality~\eqref{UB} is satisfied, and hence, $S$-matrix tree-unitarity holds.
Since the $\cO{|\B p|^4}$ contributions correspond to the leading order in the UV limit, 
these amplitudes are equivalent to those in the massless limit. 
Therefore, the $\cO{|\B p|^4}$ contributions are independent of the masses of the final states. 
This independence implies that all amplitudes at this order are identical, regardless of whether the final states consist of $F_1$ or $F_2$. 
Taking into account the factor $(-1)^{n_{IJ}}$ associated with the norm of the final states, we find that the $\cO{|\B p|^4}$ contributions cancel trivially in the sum on the right-hand side of Eq.\eqref{UB}.
Nevertheless, in Appendix~\ref{ApCal}, we explicitly evaluate each amplitude for the $(+,-)$-spins initial state, 
with which we can directly confirm this cancellation.

\section{Summary}\label{Sec:Sum}

We have demonstrated that $S$-matrix tree-unitarity,
i.e., the consistency of tree-level amplitudes with the $S$-matrix unitarity relation Eq.~\eqref{UB} in the UV limit,
holds in a fermion theory with a second-order derivative kinetic term and a power-counting renormalizable scalar-type four-Fermi interaction.
Conventional tree-unitarity~\cite{LlewellynSmith:1973yud,Cornwall:1973tb,Cornwall:1974km} implies that
the conditions for renormalizability coincide with those of the tree-level unitarity bound in the UV limit.
Although there is no rigorous proof yet, most known theories satisfy this property, such as the Glashow-Weinberg-Salam theory~\cite{tHooft:1971qjg,LlewellynSmith:1973yud}.
This is a powerful way to investigate the UV structures of theories, since it does not require loop calculations.
The unitarity bound is a necessary condition for unitarity, which is composed of $S$-matrix unitarity ($SS^\dagger=1$) and norm positivity. 
However, theories with higher-derivative (HD) kinetic term violate the second relation. 
For instance, quadratic gravity is renormalizable~\cite{Stelle:1976gc}, but it violates unitarity due to negative norm ghost excitations. 
We introduced $S$-matrix tree-unitarity in Ref.\cite{Abe:2018rwb}, which implies that renormalizability is related to one of the unitarity conditions,
namely $S$-matrix unitarity, at tree level. 
We have demonstrated that $S$-matrix tree-unitarity holds for a HD scalar theory~\cite{Abe:2018rwb} and for quadratic gravity~\cite{Abe:2022spe}. 
In this paper, we have shown that the $S$-matrix tree-unitarity is satisfied in a HD fermion theory with the scalar-type four-Fermi interaction.

A fermion with the second-order derivative kinetic term can be decomposed into two spinors with the first-order derivative, 
which was done in Sec.\ref{Sec:Dec}. 
One of them is the canonical spinor field $\psi_1$, whereas the other becomes a ghost field $\psi_2$. 
This decomposition does not change in the presence of the interaction terms,
and the interaction in terms of $\psi_1$ and $\psi_2$ can be obtained just by substituting the transformation Eq.~\eqref{FTPsi} into the original Lagrangian \eqref{HDLagWint}. 
The resulting Lagrangian is Eq.~\eqref{Lag12}. 
The canonical quantization was carried out in Sec.~\ref{sec:canonical_quantization}. 
It was shown that negative norm states for the ghost field $\psi_2$ are inevitable. 
This result differs from that of the bosonic ghost field, where, if the creation operator is associated with the negative energy modes, 
its particle states have positive norm, although perturbation theory breaks down due to the negative energy. 
Then, in Sec.~\ref{Sec:TL}, we showed that the $S$-matrix tree-unitarity holds for two-to-two scattering.

Now, having established the quantization of HD theories for both bosonic and fermionic fields, 
it would be interesting to construct supersymmetric quantum field theories.
While supersymmetric quantum field theory has been studied extensively,
it remains unclear whether it is realized in nature.
It is therefore worthwhile to explore new classes of supersymmetric quantum field theories.
In particular, combining the HD scalar theory developed in Ref.~\cite{Abe:2018rwb} with
the HD fermion theory presented here may provide a foundation for constructing a higher-derivative supersymmetric field theory.
Such a framework could open a new direction for the study of higher-derivative quantum field theories.
Furthermore, ghost-fermion actions have previously been discussed in conformal supergravity (SUGRA) \cite{Ferrara:1977mv},
although the fermion sector of \cite{Ferrara:1977mv} has third-order derivatives and the fermionic ghost sector of conformal supergravity is more complex than ours.
It would be interesting to investigate whether an interacting higher-derivative fermion theory with renormalizable interactions can also be constructed based on a third-order derivative kinetic term.

If one wants to evaluate an effective action of ghost fields, one needs a path-integral quantization.
Arisue et al.\ proposed how to write the path integral for the ghost field in the case of scalar field and gravity~\cite{Arisue:1980hk}.
See also recent discussions by Sakoda and Suzuki~\cite{Sakoda:2019vhd} and Kugo~\cite{Kugo:2026}.
It is an immediate question whether one can write the Grassmann path integral for the fermionic ghost as appearing in the HD fermion theory we have studied in this paper.

\section*{Acknowledgment}
K.~I. is supported by Japan Society for the Promotion of Science (JSPS) Grant No. JP24K07046.
Y.~A. is supported by JSPS Grants-in-Aid for Early-Career Scientists (Grant No. JP23K13108).

\appendix

\section{Feynman rules} \label{ApFeynRule}
In this appendix, we summarize the Feynman rules used in the calculation of the two-to-two scattering amplitudes.
We first specify the correspondence between the external spinors and the associated fermion lines.
We then present the four-fermion vertices derived from the scalar-type interaction in Eq.~\eqref{Lags}.

\subsection{External particle and antiparticle lines}
In this subsection, we specify the explicit forms of the Dirac spinors appearing in the mode expansion of the fermion fields introduced in Sec.~\ref{sec:canonical_quantization},
and summarize their correspondence to the external particle and antiparticle lines used in the scattering amplitudes.

In the Weyl representation, the Dirac spinor solutions are constructed from the two-component helicity eigen-spinors $\xi_s({\bf p})$ and $\eta_s({\bf p})$, which satisfy
\begin{align}
(\boldsymbol{\sigma}\cdot {\bf p})\,\xi_s({\bf p}) = s|{\bf p}|\,\xi_s({\bf p}),\qquad
(\boldsymbol{\sigma}\cdot {\bf p})\,\eta_s({\bf p})= -s|{\bf p}|\,\eta_s({\bf p}),
\end{align}
where $s=\pm1$ denotes the helicity, and
\begin{align}
\eta_s({\bf p}) = -is\sigma^2 \xi_s^*({\bf p}).
\end{align}
The particle and antiparticle spinors appearing in the mode expansion of the fermion fields introduced in Sec.~\ref{sec:canonical_quantization} are explicitly given by
\begin{align}
u_I({\bf p}, s) &=\left(\sqrt{E_{p,I} - s\left|{\bf p}\right|}\,\xi_s({\bf p}),\quad\sqrt{E_{p,I} + s\left|{\bf p}\right|}\,\xi_s({\bf p})\right)^T~~\,:=
\begin{tikzpicture} [baseline=0cm] 
\begin{feynhand}    
\setlength{\feynhanddotsize}{1mm}
\setlength{\feynhandlinesize}{1pt}
\setlength{\feynhandarrowsize}{6pt}
    \vertex [particle] (i1) at (-1.5, 0) {$F_I$};
    \vertex [dot] (f1) at (1.5, 0) {};
    \propagator [fermion] (i1) to [edge label= ${\bf p}\rightarrow$] (f1);
\end{feynhand}
\end{tikzpicture}~,\\
\overline{u}_I({\bf p}, s) &=\left(\sqrt{E_{p,I} + s\left|{\bf p}\right|}\,\xi_s^\dagger({\bf p}),\quad\sqrt{E_{p,I} - s\left|{\bf p}\right|}\,\xi_s^\dagger({\bf p})\right)~~~~:= ~~~~
\begin{tikzpicture} [baseline=0cm] 
\begin{feynhand}    
\setlength{\feynhanddotsize}{1mm}
\setlength{\feynhandlinesize}{1pt}
\setlength{\feynhandarrowsize}{6pt}
    \vertex [dot] (i1) at (-1.5, 0) {};
    \vertex [particle] (f1) at (1.5, 0) {$F_I$};
    \propagator [fermion] (i1) to [edge label= ${\bf p}\rightarrow$] (f1);
\end{feynhand}
\end{tikzpicture}~,\\
v_I({\bf p}, s)&=\left(\sqrt{E_{p,I}+s\left|{\bf p}\right|}\,\eta_s({\bf p}),\quad-\sqrt{E_{p,I}-s\left|{\bf p}\right|}\,\eta_s({\bf p})\right)^T:=\,~~~
\begin{tikzpicture} [baseline=0cm] 
\begin{feynhand}    
\setlength{\feynhanddotsize}{1mm}
\setlength{\feynhandlinesize}{1pt}
\setlength{\feynhandarrowsize}{6pt}
    \vertex [dot] (i1) at (-1.5, 0) {};
    \vertex [particle] (f1) at (1.5, 0) {$\overline{F}_I$};
    \propagator [anti fermion] (i1) to [edge label= ${\bf p}\rightarrow$] (f1);
\end{feynhand}
\end{tikzpicture}~,\\
\overline{v}_I({\bf p}, s)&=\left(-\sqrt{E_{p,I}-s\left|{\bf p}\right|}\,\eta_s^{\dag}({\bf p}),\quad\sqrt{E_{p,I}+s\left|{\bf p}\right|}\,\eta_s^{\dag}({\bf p})\right)~\,:= 
\begin{tikzpicture} [baseline=0cm] 
\begin{feynhand}    
\setlength{\feynhanddotsize}{1mm}
\setlength{\feynhandlinesize}{1pt}
\setlength{\feynhandarrowsize}{6pt}
    \vertex [particle] (i1) at (-1.5, 0) {$\overline{F}_I$};
    \vertex [dot] (f1) at (1.5, 0) {};
    \propagator [anti fermion] (i1) to [edge label= ${\bf p}\rightarrow$] (f1);
\end{feynhand}
\end{tikzpicture}~,
\end{align}
where 
\begin{align}
E_{p,I}\equiv\sqrt{|{\bf p}|^2+m_I^2}
\end{align}
is the on-shell energy of $F_I$ or $\overline{F}_I$ with mass $m_I$ and momentum ${\bf p}$.
The symbol $:=$ indicates the correspondence between each external particle or antiparticle state, represented by its Dirac spinor, and the associated external line in the Feynman diagrams.

\subsection{Four-fermion vertices}
In this subsection, we derive the four-fermion vertex functions from the scalar-type interaction in Eq.~\eqref{Lags}.
The interaction Lagrangian is given by
\begin{align}
\mathcal{L}_s &= \frac{\lambda_s}{2(m_2-m_1)^2} \left( \sum_{I,J} \epsilon_I\epsilon_J \overline{\psi}_I\psi_J \right)^2
=\frac{\lambda_s}{2(m_2-m_1)^2} \sum_{I,J,K,L} \epsilon_I\epsilon_J\epsilon_K\epsilon_L \left( \overline{\psi}_I\psi_J \right)\left( \overline{\psi}_K\psi_L \right).
\label{Lags_expand_4th-Fermi}
\end{align}
Although Eq.~\eqref{Lags_expand_4th-Fermi} is written in terms of the fermion fields $\psi_I$ and $\overline{\psi}_I$,
the corresponding Feynman rules are more conveniently expressed in terms of the particle and antiparticle external states.
Below, we summarize the vertex functions corresponding to the external states $F_I F_J F_K F_L$, $F_I \overline{F}_J F_K\overline{F}_L$, and $\overline{F}_I\overline{F}_J\overline{F}_K\overline{F}_L$.
\begin{align}
\begin{tikzpicture} [baseline=0cm] 
\begin{feynhand}    
\setlength{\feynhanddotsize}{2mm}
\setlength{\feynhandlinesize}{1pt}
\setlength{\feynhandarrowsize}{6pt}
    \vertex [particle] (i1) at (-0.7,-0.7) {$F_I$};
    \vertex [particle] (i2) at (0.7,-0.7) {$F_J$};
    \vertex [particle] (f1) at (-0.7,0.7) {$F_K$};
    \vertex [particle] (f2) at (0.7,0.7) {$F_L$};
    \vertex (a) at (0,0); 
    \propagator [plain] (i1) to [] (a);
    \propagator [plain] (i2) to [] (a);
    \propagator [plain] (a) to [] (f1);
    \propagator [plain] (a) to [] (f2);
\end{feynhand}
\end{tikzpicture}
=
\begin{tikzpicture} [baseline=0cm] 
\begin{feynhand}    
\setlength{\feynhanddotsize}{2mm}
\setlength{\feynhandlinesize}{1pt}
\setlength{\feynhandarrowsize}{6pt}
    \vertex [particle] (i1) at (-0.7,-0.7) {$F_I$};
    \vertex [particle] (i2) at (0.7,-0.7) {$\overline{F}_J$};
    \vertex [particle] (f1) at (-0.7,0.7) {$F_K$};
    \vertex [particle] (f2) at (0.7,0.7) {$\overline{F}_L$};
    \vertex (a) at (0,0); 
    \propagator [plain] (i1) to [] (a);
    \propagator [plain] (i2) to [] (a);
    \propagator [plain] (a) to [] (f1);
    \propagator [plain] (a) to [] (f2);
\end{feynhand}
\end{tikzpicture}
=
\begin{tikzpicture} [baseline=0cm] 
\begin{feynhand}    
\setlength{\feynhanddotsize}{2mm}
\setlength{\feynhandlinesize}{1pt}
\setlength{\feynhandarrowsize}{6pt}
    \vertex [particle] (i1) at (-0.7,-0.7) {$\overline{F}_I$};
    \vertex [particle] (i2) at (0.7,-0.7) {$\overline{F}_J$};
    \vertex [particle] (f1) at (-0.7,0.7) {$\overline{F}_K$};
    \vertex [particle] (f2) at (0.7,0.7) {$\overline{F}_L$};
    \vertex (a) at (0,0); 
    \propagator [plain] (i1) to [] (a);
    \propagator [plain] (i2) to [] (a);
    \propagator [plain] (a) to [] (f1);
    \propagator [plain] (a) to [] (f2);
\end{feynhand}
\end{tikzpicture}
\qquad&:=\epsilon_I\epsilon_J\epsilon_K\epsilon_L\frac{\lambda_s}{(m_2-m_1)^2},
\end{align}
where $\epsilon_1=1$ and $\epsilon_2=-1$.
We note that the sign of each vertex depends on the number of ghost fermions attached to its external lines.
This sign dependence is represented by the product of the factors $\epsilon_I\epsilon_J\epsilon_K\epsilon_L$.
In particular, a vertex with an odd number of ghost-fermion external lines has the opposite sign to one with an even number of ghost-fermion external lines.


\section{Detailed Calculation of Scattering Amplitudes} \label{ApCal}

In this appendix, we present the calculation of the amplitudes for two-to-two scattering processes.
The interaction term under consideration is restricted to the scalar-type shown in Eq.~\eqref{Lags}. 
The lowest-order contributions to the four-point functions are described by tree-level diagrams.
Using the Feynman rules derived in Appendix~\ref{ApFeynRule}, the amplitudes for the two-to-two scattering processes are represented by the following contact diagrams.
\begin{align}
&\begin{tikzpicture} [baseline=0cm] 
\begin{feynhand}    
\setlength{\feynhanddotsize}{2mm}
\setlength{\feynhandlinesize}{1pt}
\setlength{\feynhandarrowsize}{6pt}
    \vertex [particle] (i1) at (-1.5,-1.5) {$F_1$};
    \vertex [particle] (i2) at (1.5,-1.5) {$F_1$};
    \vertex [particle] (f1) at (-1.5,1.5) {$F_I$};
    \vertex [particle] (f2) at (1.5,1.5) {$F_J$};
    \vertex (a) at (0,0); 
    \propagator [fermion] (i1) to [edge label= $\B p$] (a);
    \propagator [fermion] (i2) to [edge label'=$-\B p$] (a);
    \propagator [fermion] (a) to [edge label= $\B q_{IJ}$] (f1);
    \propagator [fermion] (a) to [edge label'= $-\B q_{IJ}$] (f2);
\end{feynhand}
\end{tikzpicture}
\qquad=i\mathcal{M}\bigl(F_{1}(\B p,+)F_{1}(-\B p,\pm)\rightarrow F_{I}(\B q_{IJ},s)F_{J}(-\B q_{IJ},s')\bigr)\nonumber\\
&= i\frac{\lambda_s \epsilon_{I}\epsilon_{J} }{(m_2-m_1)^2} \left[\bar{u}_I(\mathbf{q}_{IJ}, s) u_1(\mathbf{p}, +)  \bar{u}_J(-\mathbf{q}_{IJ}, s') u_1(-\mathbf{p}, \pm) 
- \bar{u}_J(-\mathbf{q}_{IJ}, s') u_1(\mathbf{p}, +)   \bar{u}_I(\mathbf{q}_{IJ}, s) u_1(-\mathbf{p}, \pm)\right]. 
\label{two-body scattering of F-F-F-F}
\end{align}

The spinor bilinear appearing in Eq.~\eqref{two-body scattering of F-F-F-F} is written as
\begin{eqnarray}
\bar{u}_J(\mathbf{q}, s') u_I(\mathbf{p}, s) 
= \left( \sqrt{E_{\B q,J}+s|\mathbf{q}|}\sqrt{E_{\B p,I}-s'|\mathbf{p}|} + \sqrt{E_{\B q,J}-s|\mathbf{q}|}\sqrt{E_{\B p,I}+s'|\mathbf{p}|} \right) \xi_{s'}^\dagger(\mathbf{q}) \xi_s(\mathbf{p}).
\label{spinor_bilinear}
\end{eqnarray}
The inner product of the two-component helicity eigenspinors is
\begin{eqnarray}
\xi_{s'}^{\dag}(\mathbf q)\xi_{s}(\mathbf p) &=& \frac{1+ss'}{2\sqrt{2}} \sqrt{1 + \frac{ \mathbf{p} \cdot \mathbf{q}}{|\mathbf{p}||\mathbf{q}|}} - s \frac{1-ss'}{2\sqrt{2}} \sqrt{1 - \frac{ \mathbf{p} \cdot \mathbf{q}}{|\mathbf{p}||\mathbf{q}|}} \, \nn 
&=& \frac{1+ss'}{2} \cos \frac{\theta}{2} - s \frac{1-ss'}{2} \sin \frac{\theta}{2} ,
\end{eqnarray}
where $\theta$ is the angle between $\mathbf{p}$ and $\mathbf{q}$. For other pairs of momenta in the
scattering plane, the second expression holds with $\theta$ replaced by the oriented angle from the
momentum of $\xi_s$ to that of $\xi^{\dagger}_{s'}$, with $\mathbf{p}$, $\mathbf{q}$, $-\mathbf{p}$ and
$-\mathbf{q}$ placed at the angles $0$, $\theta$, $\pi$ and $\pi+\theta$, respectively. Since spinors
change sign under a $2\pi$ rotation, this angle is defined modulo $4\pi$ and cannot be replaced by the
angle between the two vectors: for example, $\pi+\theta$ and $\theta-\pi$, rather than $\pi-\theta$,
are substituted for $\xi^{\dagger}_{s'}(-\mathbf{q})\xi_s(\mathbf{p})$ and
$\xi^{\dagger}_{s'}(\mathbf{q})\xi_s(-\mathbf{p})$, respectively.

To determine the UV behavior of the scattering amplitudes, we expand the kinematic factors appearing in Eq.~\eqref{spinor_bilinear}.
We take the UV limit $|\mathbf{p}|\to\infty$ with the scattering angle fixed.
The magnitude of the final-state momentum, $|\mathbf{q}|:=|\mathbf{q}_{IJ}|$, is expressed in terms of $|\mathbf p|$ through energy conservation.
The relevant asymptotic expansions in the UV region, $|\mathbf p|,|\mathbf q|\gg m_I,m_J$, are
\begin{eqnarray}
&&\sqrt{E_{\mathbf{p},I} + |\mathbf{p}|} = \sqrt{2|\mathbf{p}|} \left( 1 + \frac{1}{8}\frac{m_I^2}{|\mathbf{p}|^2} + \cO{\frac{m_I^4}{|\mathbf{p}|^4}} \right) , \\
&&\sqrt{E_{\mathbf{p},I} - |\mathbf{p}|} = \frac{m_I}{\sqrt{2|\mathbf{p}|}}  + \cO{\frac{m_I^3}{|\mathbf{p}|^{5/2}}} .
\end{eqnarray}
From these expansions, the products of the kinematic factors that enter the spinor bilinears are obtained as
\begin{eqnarray}
&&\sqrt{E_{\mathbf{p},I} + |\mathbf{p}|} \sqrt{E_{\mathbf{q},J} + |\mathbf{q}|} 
= 2\sqrt{|\mathbf{p}||\mathbf{q}|} \left( 1 + \frac{1}{8}\frac{m_I^2}{|\mathbf{p}|^2}+ \frac{1}{8}\frac{m_J^2}{|\mathbf{q}|^2} +\cO{\frac{m^4}{|\mathbf{p}|^4}} \right) , \\
&&\sqrt{E_{\mathbf{p},I} + |\mathbf{p}|} \sqrt{E_{\mathbf{q},J} - |\mathbf{q}|} 
= \frac{m_J\sqrt{|\mathbf{p}|}}{\sqrt{|\mathbf{q}|}}  + \cO{\frac{m^3}{|\mathbf{p}|^2}} , \\
&&\sqrt{E_{\mathbf{p},I} - |\mathbf{p}|} \sqrt{E_{\mathbf{q},J} - |\mathbf{q}|}  = \frac{m_Im_J}{2\sqrt{|\mathbf{p}||\mathbf{q}|}}  + \cO{\frac{m^4}{|\mathbf{p}|^3}} .
\end{eqnarray}
The combinations that appear directly in the spinor inner products are then given by
\begin{eqnarray}
&&\sqrt{E_{\mathbf{p},I} + |\mathbf{p}|} \sqrt{E_{\mathbf{q},J} + |\mathbf{q}|} + \sqrt{E_{\mathbf{p},I} - |\mathbf{p}|} \sqrt{E_{\mathbf{q},J} - |\mathbf{q}|}\nn 
&&\hspace{10mm}
= 2\sqrt{|\mathbf{p}||\mathbf{q}|} \left( 1 + \frac{1}{8}\frac{m_I^2}{|\mathbf{p}|^2}+ \frac{1}{8}\frac{m_J^2}{|\mathbf{q}|^2}+ \frac{1}{4}\frac{m_Im_J}{|\mathbf{p}||\mathbf{q}|} + \cO{\frac{m^4}{|\mathbf{p}|^4}} \right) , \\
&&\sqrt{E_{\mathbf{p},I} + |\mathbf{p}|}\,\sqrt{E_{\mathbf{q},J} - |\mathbf{q}|}
+\sqrt{E_{\mathbf{p},I} - |\mathbf{p}|}\,\sqrt{E_{\mathbf{q},J} + |\mathbf{q}|} 
=  
m_J \sqrt{\frac {|\mathbf p|}{|\mathbf q|}} + m_I \sqrt{\frac{|\mathbf q|}{|\mathbf p|}} 
 + \cO{\frac{m^3}{|\mathbf{p}|^2}} .
\end{eqnarray}

We consider two initial states: i) (+,+) spins, ii) $(+,-)$ spins. 
Both of the initial particles are positive norm particles $F_1$ with momenta $\B p$ and $-\B p$, 
which means that we work in the center-of-mass frame. 
We consider all the cases where the final particles are either $F_1$ or $F_2$ with arbitrary spins.
The momenta of the final particles are denoted by $\B q_{IJ}$ and $-\B q_{IJ}$. 
In each subsection, we present the calculation of the scattering amplitudes with the same set of final spins.
The results for both cases with i) (+,+) initial spins, ii) $(+,-)$ initial spins are shown in the same subsection, and are distinguished by i) and ii).

\subsection{$\M \bigl(F_1(\B p, +)F_1(-\B p, \pm) \to F_1(\B q_{11}, s)F_1(-\B q_{11}, s') \bigr)$}
Since the final state has two $F_1$ particles, which is the same as the initial state, 
the energy momentum conservation gives $|\B p|=|\B q_{11}|$.
Then, the relevant spinor bilinears, $\bar{u}_1(\pm\mathbf{q}, \pm) u_1(\pm\mathbf{p}, \pm)$, are 
\begin{eqnarray}
&&\bar{u}_1(\mathbf{q}_{11}, \pm) u_1(\mathbf{p}, \pm)  = \bar{u}_1(-\mathbf{q}_{11}, \pm) u_1(-\mathbf{p}, \pm) =2m_1 \left( 1+\cO{\frac{m^2}{|\B p|^2}} \right) \cos \frac{\theta}{2}, \\
&&-\bar{u}_1(-\mathbf{q}_{11}, \pm) u_1(\mathbf{p}, \pm)= \bar{u}_1(\mathbf{q}_{11}, \pm) u_1(-\mathbf{p}, \pm)
= 2m_1 \left( 1+\cO{\frac{m^2}{|\B p|^2}} \right) \sin \frac{\theta}{2}, \\
&&\bar{u}_1(\mathbf{q}_{11}, \pm) u_1(\mathbf{p}, \mp) = \bar{u}_1(-\mathbf{q}_{11}, \pm) u_1(-\mathbf{p}, \mp)
= \pm 2|\B p| \left( 1+ \frac{m_1^2}{2|\B p|^2}+\cO{\frac{m^4}{|\B p|^4}} \right) \sin\frac{\theta}{2}, \\
&&\bar{u}_1(-\mathbf{q}_{11}, \pm) u_1(\mathbf{p}, \mp) =-\bar{u}_1(\mathbf{q}_{11}, \pm) u_1(-\mathbf{p}, \mp)
= \pm 2|\B p| \left( 1+ \frac{m_1^2}{2|\B p|^2}+\cO{\frac{m^4}{|\B p|^4}} \right) \cos\frac{\theta}{2}, 
\end{eqnarray}
where the signs $\pm$ and $\mp$ are to be taken in the same order. 
This gives, for i) (+,+) initial spins,
\begin{eqnarray}
&&\M \bigl(F_1(\B p, +)F_1(-\B p, +) \to F_1(\B q_{11}, +)F_1(-\B q_{11}, -) \bigr) \nonumber \\
&&\hspace{6mm}
= \M \bigl(F_1(\B p, +)F_1(-\B p, +) \to F_1(\B q_{11}, -)F_1(-\B q_{11}, +) \bigr) =0 ,  \\
&& \M \bigl(F_1(\B p, +)F_1(-\B p, +) \to F_1(\B q_{11}, +)F_1(-\B q_{11}, +) \bigr) \nonumber \\
&&\hspace{6mm}
= \frac{\lambda_s}{(m_2-m_1)^2} \left[4m_1^2 \cos^2\frac{\theta}{2}+4m_1^2 \sin^2\frac{\theta}{2}  + \cO{\frac{m^4}{|\B p|^2}} \right]\nonumber \\
&&\hspace{6mm}
= \frac{\lambda_s}{(m_2-m_1)^2} \left[4m_1^2   + \cO{\frac{m^4}{|\B p|^2}} \right], \\
&& \M \bigl(F_1(\B p, +)F_1(-\B p, +) \to F_1(\B q_{11}, -)F_1(-\B q_{11}, -) \bigr) \nonumber \\
&&\hspace{6mm}
= \frac{\lambda_s}{(m_2-m_1)^2} \left[4\left(|\B p|^2 + m_1^2\right) \sin^2\frac{\theta}{2}+4\left(|\B p|^2 + m_1^2\right)\cos^2\frac{\theta}{2}  + \cO{\frac{m^4}{|\B p|^2}} \right]\nonumber \\
&&\hspace{6mm}
= \frac{\lambda_s}{(m_2-m_1)^2} \left[4\left(|\B p|^2 + m_1^2\right)  + \cO{\frac{m^4}{|\B p|^2}} \right] ,
\end{eqnarray}
and, for ii) $(+,-)$ initial spins,
\begin{eqnarray}
&&\M \bigl(F_1(\B p, +)F_1(-\B p, -) \to F_1(\B q_{11}, +)F_1(-\B q_{11}, +) \bigr) \nn
&&\hspace{6mm}
= \M \bigl(F_1(\B p, +)F_1(-\B p, -) \to F_1(\B q_{11}, -)F_1(-\B q_{11}, -) \bigr) =0 ,\\
&&\M \bigl(F_1(\B p, +)F_1(-\B p, -) \to F_1(\B q_{11}, +)F_1(-\B q_{11}, -) \bigr)  \nn
&&\hspace{6mm}
= \frac{\lambda_s}{(m_2-m_1)^2} \left[4m_1^2 \cos^2\frac{\theta}{2} - 4 (|\B p|^2 + m_1^2) \cos^2\frac{\theta}{2} + \cO{\frac{m^4}{|\B p|^2}} \right]
\nn
&&\hspace{6mm}
= \frac{4\lambda_s}{(m_2-m_1)^2} \left[- |\B p|^2 \cos^2\frac{\theta}{2}+ \cO{\frac{m^4}{|\B p|^2}}\right] ,
\label{+-+-}
\\
&& \M \bigl(F_1(\B p, +)F_1(-\B p, -) \to F_1(\B q_{11}, -)F_1(-\B q_{11}, +) \bigr) \nonumber \\
&&\hspace{6mm}
= \frac{\lambda_s}{(m_2-m_1)^2} \left[4m_1^2 \sin^2\frac{\theta}{2} - 4 (|\B p|^2 + m_1^2) \sin^2\frac{\theta}{2} + \cO{\frac{m^4}{|\B p|^2}} \right] 
\nn
&&\hspace{6mm}
= \frac{4\lambda_s}{(m_2-m_1)^2} \left[-|\B p|^2 \sin^2\frac{\theta}{2}+ \cO{\frac{m^4}{|\B p|^2}}\right] .
\end{eqnarray}

\subsection{$\M \bigl(F_1(\B p, +)F_1(-\B p, \pm) \to F_1(\B q_{12}, s)F_2(-\B q_{12}, s') \bigr)$}
Since the mass of one of the final-state particles differs from that of the initial-state particles,
the magnitude of the final momentum, $|\B q_{12}|$, differs from the initial one, $|\B p|$.
This relation is obtained from the energy conservation
\begin{eqnarray}
2\sqrt{|\B p|^2+m_1^2} = \sqrt{|\B q_{12}|^2+m_1^2}+\sqrt{|\B q_{12}|^2+m_2^2}, 
\end{eqnarray}
which yields 
\begin{eqnarray}
|\B q_{12}|= |\B p|\left[1 + \frac{m_1^2-m_2^2}{4|\B p|^2}+ \cO{\frac{m^4}{|\B p|^4}} \right].
\end{eqnarray}
This gives
\begin{eqnarray}
&&\sqrt{E_{\mathbf{q}_{12},I} + |\mathbf{q}_{12}|} = \sqrt{2|\mathbf{p}|} \left( 1 + \frac{m_I^2+m_1^2-m_2^2}{8|\mathbf{p}|^2} + \cO{\frac{m_I^4}{|\mathbf{p}|^4}} \right) , \\
&&\sqrt{E_{\mathbf{q}_{12},I} - |\mathbf{q}_{12}|} = \frac{m_I}{\sqrt{2|\mathbf{p}|}}  + \cO{\frac{m_I^3}{|\mathbf{p}|^{5/2}}}.
\end{eqnarray}

By using the above equations, the spinor bilinears, $\bar{u}_I(\pm\mathbf{q}_{12}, \pm) u_1(\pm\mathbf{p}, \pm)$, are calculated, 
and the results become
\begin{align}
&\bar{u}_1(\mathbf{q}_{12}, \pm) u_1(\mathbf{p}, \pm)= \bar{u}_1(-\mathbf{q}_{12}, \pm) u_1(-\mathbf{p}, \pm)
= 2 m_1\left(1  +\cO{\frac{m^2}{|\B p|^2}} \right) \cos\frac{\theta}{2} ,  \\
&-\bar{u}_1(-\mathbf{q}_{12}, \pm) u_1(\mathbf{p}, \pm)= \bar{u}_1(\mathbf{q}_{12}, \pm) u_1(-\mathbf{p}, \pm)
= 2 m_1\left(1  +\cO{\frac{m^2}{|\B p|^2}} \right) \sin\frac{\theta}{2} ,  \\
&\bar{u}_1(\mathbf{q}_{12}, \pm) u_1(\mathbf{p}, \mp)=\bar{u}_1(-\mathbf{q}_{12}, \pm) u_1(-\mathbf{p}, \mp) 
= \pm 2|\B p| \left( 1+ \frac{5 m_1^2-m_2^2}{8|\B p|^2}
+\cO{\frac{m^4}{|\B p|^4}} \right) \sin\frac{\theta}{2}, \\
&\bar{u}_1(-\mathbf{q}_{12}, \pm) u_1(\mathbf{p}, \mp) =-\bar{u}_1(\mathbf{q}_{12}, \pm) u_1(-\mathbf{p}, \mp)
= \pm 2|\B p| \left( 1+ \frac{5 m_1^2-m_2^2}{8|\B p|^2}
+\cO{\frac{m^4}{|\B p|^4}} \right) \cos \frac{\theta}{2} , \\
&\bar{u}_2(\mathbf{q}_{12}, \pm) u_1(\mathbf{p}, \pm)= \bar{u}_2(-\mathbf{q}_{12}, \pm) u_1(-\mathbf{p}, \pm)
= (m_1+m_2)\left(1  +\cO{\frac{m^2}{|\B p|^2}} \right) \cos\frac{\theta}{2},  \\
&-\bar{u}_2(-\mathbf{q}_{12}, \pm) u_1(\mathbf{p}, \pm)= \bar{u}_2(\mathbf{q}_{12}, \pm) u_1(-\mathbf{p}, \pm)
= (m_1+m_2)\left(1  +\cO{\frac{m^2}{|\B p|^2}} \right) \sin\frac{\theta}{2},  \\
&\bar{u}_2(\mathbf{q}_{12}, \pm) u_1(\mathbf{p}, \mp) =\bar{u}_2(-\mathbf{q}_{12}, \pm) u_1(-\mathbf{p}, \mp) 
= \pm 2|\B p|  \left[ 1 + \frac{m_1(m_1 + m_2)}{4|\mathbf{p}|^2} 
+\cO{\frac{m^4}{|\B p|^4}} \right] \sin\frac{\theta}{2},\\
&\bar{u}_2(-\mathbf{q}_{12}, \pm) u_1(\mathbf{p}, \mp) =-\bar{u}_2(\mathbf{q}_{12}, \pm) u_1(-\mathbf{p}, \mp) 
= \pm 2|\B p|  \left[ 1 + \frac{m_1(m_1 + m_2)}{4|\mathbf{p}|^2} 
+\cO{\frac{m^4}{|\B p|^4}} \right] \cos \frac{\theta}{2}.
\end{align}

The scattering amplitudes are obtained as, for i) (+,+) initial spins,
\begin{eqnarray}
&&\M \bigl(F_1(\B p, +)F_1(-\B p, +) \to F_1(\B q_{12}, +)F_2(-\B q_{12}, +) \bigr) 
\nn
&&\hspace{6mm}
= \frac{ - 2 \lambda_s}{(m_2-m_1)^2}  \left(m_1(m_1+m_2)  + \cO{\frac{m^4}{|\B p|^2}}\right),\\
&&\M \bigl(F_1(\B p, +)F_1(-\B p, +) \to F_1(\B q_{12}, -)F_2(-\B q_{12}, -) \bigr) 
\nn
&&\hspace{6mm}
= \frac{ - \lambda_s}{(m_2-m_1)^2}  \left(4|\B p|^2+ \frac{7m_1^2+2m_1m_2-m_2^2}{2}  + \cO{\frac{m^4}{|\B p|^2}}\right),
\\
&&\M \bigl(F_1(\B p, +)F_1(-\B p, +) \to F_1(\B q_{12}, +)F_2(-\B q_{12}, -) \bigr) 
\nn
&&\hspace{6mm}
= \M \bigl(F_1(\B p, +)F_1(-\B p, +) \to F_1(\B q_{12}, -)F_2(-\B q_{12}, +) \bigr)  
=0, 
\end{eqnarray}
and, for ii) $(+,-)$ initial spins,
\begin{eqnarray}
&&\M \bigl(F_1(\B p, +)F_1(-\B p, -) \to F_1(\B q_{12}, +)F_2(-\B q_{12}, +) \bigr) = \M \bigl(F_1(\B p, +)F_1(-\B p, -) \to F_1(\B q_{12}, -)F_2(-\B q_{12}, -) \bigr)  \nn
&&\hspace{6mm}
= \frac{-2 \lambda_s}{(m_2-m_1)^2} \left( |\B p| (m_1-m_2) \cos \frac{\theta}{2} \sin \frac{\theta}{2}+ \cO{\frac{m^3}{|\B p|}} \right),
\\
&&\M \bigl(F_1(\B p, +)F_1(-\B p, -) \to F_1(\B q_{12}, +)F_2(-\B q_{12}, -) \bigr) 
\nn
&&\hspace{6mm}
=  \frac{\lambda_s}{(m_2-m_1)^2} \left(  \left[4|\B p|^2 +\frac{3m_1^2-2m_1m_2-m_2^2}{2}\right]\cos^2\frac{\theta}{2}
+ \cO{\frac{m^4}{|\B p|^2}} \right), \\
&& \M \bigl(F_1(\B p, +)F_1(-\B p, -) \to F_1(\B q_{12}, -)F_2(-\B q_{12}, +) \bigr)  
\nn
&&\hspace{6mm}
=  \frac{\lambda_s}{(m_2-m_1)^2}  \left(\left[4|\B p|^2 +\frac{3m_1^2-2m_1m_2-m_2^2}{2}\right]\sin^2\frac{\theta}{2}
+ \cO{\frac{m^4}{|\B p|^2}} \right).
\end{eqnarray}

\subsection{$\M \bigl(F_1(\B p, +)F_1(-\B p, \pm) \to F_2(\B q_{22}, s)F_2(-\B q_{22}, s') \bigr)$}
The relation between $|\B q_{22}|$ and $|\B p|$ is obtained from the energy conservation
\begin{eqnarray}
2\sqrt{|\B p|^2+m_1^2} = 2\sqrt{|\B q_{22}|^2+m_2^2}. 
\end{eqnarray}
This gives 
\begin{eqnarray}
|\B q_{22}|= |\B p|\left[1 + \frac{m_1^2-m_2^2}{2|\B p|^2} + \cO{\frac{m^4}{|\B p|^4}} \right],
\end{eqnarray}
and we have 
\begin{eqnarray}
&&\sqrt{E_{\mathbf{q}_{22},I} + |\mathbf{q}_{22}|} = \sqrt{2|\mathbf{p}|} \left( 1 + \frac{m_I^2+2(m_1^2-m_2^2)}{8|\mathbf{p}|^2} + \cO{\frac{m_I^4}{|\mathbf{p}|^4}} \right) , \\
&&\sqrt{E_{\mathbf{q}_{22},I} - |\mathbf{q}_{22}|} = \frac{m_I}{\sqrt{2|\mathbf{p}|}}  + \cO{\frac{m_I^3}{|\mathbf{p}|^{5/2}}} .
\end{eqnarray}

The spinor bilinears, $\bar{u}_2(\pm\mathbf{q}_{22}, \pm) u_1(\pm\mathbf{p}, \pm)$, are calculated as
\begin{eqnarray}
&&\bar{u}_2(\mathbf{q}_{22}, \pm) u_1(\mathbf{p}, \pm)=\bar{u}_2(-\mathbf{q}_{22}, \pm) u_1(-\mathbf{p}, \pm)
= (m_1 + m_2) \left[1 +\cO{\frac{m^2}{|\B p|^2}}\right] \cos\frac{\theta}{2},  \\
&&-\bar{u}_2(-\mathbf{q}_{22}, \pm) u_1(\mathbf{p}, \pm)=\bar{u}_2(\mathbf{q}_{22}, \pm) u_1(-\mathbf{p}, \pm)
= (m_1 + m_2) \left[1 +\cO{\frac{m^2}{|\B p|^2}}\right] \sin\frac{\theta}{2},  \\
&&\bar{u}_2(\mathbf{q}_{22}, \pm) u_1(\mathbf{p}, \mp) =\bar{u}_2(-\mathbf{q}_{22}, \pm) u_1(-\mathbf{p}, \mp) \nn
&&\hspace{6mm}
= \pm 2|\B p| \left( 1+ \frac{3m_1^2 + 2m_1m_2 - m_2^2}{8|\mathbf{p}|^2}
+\cO{\frac{m^4}{|\B p|^4}} \right) \sin\frac{\theta}{2} ,\\
&&\bar{u}_2(-\mathbf{q}_{22}, \pm) u_1(\mathbf{p}, \mp) =-\bar{u}_2(\mathbf{q}_{22}, \pm) u_1(-\mathbf{p}, \mp)\nn
&&\hspace{6mm} 
= \pm 2|\B p| \left( 1+ \frac{3m_1^2 + 2m_1m_2 - m_2^2}{8|\mathbf{p}|^2}
+\cO{\frac{m^4}{|\B p|^4}} \right) \cos\frac{\theta}{2}.
\end{eqnarray}

Then, the scattering amplitudes are obtained as, for i) (+,+) initial spins,
\begin{eqnarray}
&&\M \bigl(F_1(\B p, +)F_1(-\B p, +) \to F_2(\B q_{22}, +)F_2(-\B q_{22}, +) \bigr) 
\nn
&&\hspace{6mm}
= \frac{\lambda_s}{(m_2-m_1)^2} \left((m_1+m_2)^2
+\cO{\frac{m^4}{|\B p|^2}}\right) ,\\
&& \M \bigl(F_1(\B p, +)F_1(-\B p, +) \to F_2(\B q_{22}, -)F_2(-\B q_{22}, -) \bigr) 
\nn
&&\hspace{6mm}
= \frac{  \lambda_s}{(m_2-m_1)^2} \left(4|\B p|^2 +3m_1^2+2m_1m_2-m_2^2
+\cO{\frac{m^4}{|\B p|^2}}\right) , \\
&&\M \bigl(F_1(\B p, +)F_1(-\B p, +) \to F_2(\B q_{22}, +)F_2(-\B q_{22}, -) \bigr) 
\nn
&&\hspace{6mm}=\M \bigl(F_1(\B p, +)F_1(-\B p, +) \to F_2(\B q_{22}, -)F_2(-\B q_{22}, +) \bigr)  
=0 ,
\end{eqnarray}
and, for ii) $(+,-)$ initial spins,
\begin{eqnarray}
&&\M \bigl(F_1(\B p, +)F_1(-\B p, -) \to F_2(\B q_{22}, +)F_2(-\B q_{22}, +) \bigr) 
\nn
&&\hspace{6mm}
= \M \bigl(F_1(\B p, +)F_1(-\B p, -) \to F_2(\B q_{22}, -)F_2(-\B q_{22}, -) \bigr) =0 ,  \\
&&\M \bigl(F_1(\B p, +)F_1(-\B p, -) \to F_2(\B q_{22}, +)F_2(-\B q_{22}, -) \bigr) 
\nn
&&\hspace{6mm}
= \frac{-\lambda_s}{(m_2-m_1)^2} \left(\left[ 4|\B p|^2 +2(m_1^2-m_2^2)\right]\cos^2 \frac{\theta}{2}
+\cO{\frac{m^4}{|\B p|^2}}\right), \\
&&\M \bigl(F_1(\B p, +)F_1(-\B p, -) \to F_2(\B q_{22}, -)F_2(-\B q_{22}, +) \bigr)  
\nn
&&\hspace{6mm}
=  \frac{-\lambda_s}{(m_2-m_1)^2}\left(\left[ 4|\B p|^2 +2 (m_1^2-m_2^2)\right]\sin^2 \frac{\theta}{2}
+\cO{\frac{m^4}{|\B p|^2}} \right) .
\end{eqnarray}


\bibliography{HDFermionRef.bib}
\bibliographystyle{unsrt}

\end{document}